\documentclass[8.5pt,twoside,twocolumn]{article}
\usepackage[super,sort&compress,comma]{natbib} 
\usepackage{mhchem}
\usepackage{times,mathptmx}
\usepackage{sectsty}
\usepackage{balance} 

\usepackage{graphicx} 
\usepackage{lastpage}
\usepackage[format=plain,justification=raggedright,singlelinecheck=false,font=small,labelfont=bf,labelsep=space]{caption} 
\usepackage{fancyhdr}
\usepackage{}

\renewcommand{\vec}[1]{{\bf #1}}
\newcommand \be {\begin{equation}}
\newcommand \ee {\end{equation}}
\newcommand \bea {\begin{eqnarray}}
\newcommand \eea {\end{eqnarray}}

\begin{document}

\def\NOTE#1{{\textcolor{blue}{ [#1]}}}  
\def\REPLACE#1{{\textcolor{red}{ #1}}}  

\thispagestyle{plain}
\fancypagestyle{plain}{
\fancyhead[L]{Advance Article}
\fancyhead[R]{http://xlink.rsc.org/?doi=C2SM07090A\vspace{-0.2cm}}
\renewcommand{\headrulewidth}{1pt}}
\renewcommand{\thefootnote}{\fnsymbol{footnote}}
\renewcommand\footnoterule{\vspace*{1pt}%
\hrule width 3.4in height 0.4pt \vspace*{5pt}} 
\setcounter{secnumdepth}{5}

\makeatletter 
\def\subsubsection{\@startsection{subsubsection}{3}{10pt}{-1.25ex plus -1ex minus -.1ex}{0ex plus 0ex}{\normalsize\bf}} 
\def\paragraph{\@startsection{paragraph}{4}{10pt}{-1.25ex plus -1ex minus -.1ex}{0ex plus 0ex}{\normalsize\textit}} 
\renewcommand\@biblabel[1]{#1}            
\renewcommand\@makefntext[1]%
{\noindent\makebox[0pt][r]{\@thefnmark\,}#1}
\makeatother 
\renewcommand{\figurename}{\small{Fig.}~}
\sectionfont{\large}
\subsectionfont{\normalsize} 

\fancyfoot{}
\fancyfoot[CO]{\vspace{-7.2pt}\hspace{12.2cm} Soft Matter (2012)}
\fancyfoot[CE]{\vspace{-7.5pt}\hspace{-13.5cm} Soft Matter (2012)}
\fancyfoot[RO]{\footnotesize{\sffamily{1--\pageref{LastPage} ~\textbar  \hspace{2pt}\thepage}}}
\fancyfoot[LE]{\footnotesize{\sffamily{\thepage~\textbar\hspace{3.45cm} 1--\pageref{LastPage}}}}
\fancyhead{}
\renewcommand{\headrulewidth}{1pt} 
\renewcommand{\footrulewidth}{1pt}
\setlength{\arrayrulewidth}{1pt}
\setlength{\columnsep}{6.5mm}
\setlength\bibsep{1pt}

\twocolumn[
  \begin{@twocolumnfalse}
\noindent\LARGE{\textbf{Spontaneous formation of permanent shear bands in a mesoscopic model of flowing disordered matter}}
\vspace{0.6cm}

\noindent\large{\textbf{Kirsten Martens,\textit{$^{a*}$} Lyd\'eric Bocquet,\textit{$^{a}$} and
Jean-Louis Barrat\textit{$^{b}$}}}\vspace{0.5cm}
%
%

\noindent \textbf{\small{DOI: 10.1039/C2SM07090A}}
\vspace{0.6cm}

\noindent \normalsize{This study proposes a coherent scenario of the formation of permanent shear bands in the flow of yield stress materials. It is a well accepted point of view that flow in disordered media is occurring via local plastic events, corresponding to small size rearrangements, that yield a long range stress redistribution over the system. Within a minimalistic mesoscopic model that incorporates these local dynamics, we study the spatial organisation of the local plastic events. The most important parameter in this study is the typical restructuring time needed to regain the original structure after a local rearrangement.
In agreement with a recent mean field study [Coussot \textit{et al., Eur.~Phys.~J.~E}, 2010, \textbf{33}, 183] we observe a spontaneous formation of permanent shear bands, when this restructuring time is large compared to the typical stress release time  in a rearrangement. The bands consist of a large number of plastic events within a solid region that remains elastic. This heterogeneous flow behaviour is different in nature from the transient dynamical heterogeneities that one observes in the small shear rate limit in flow without shear-banding [Martens \textit{et al., Phys.~Rev.~Lett.}, 2011, \textbf{106}, 156001]. We analyse in detail the dependence of the shear bands on system size, shear rate and restructuring time. Further we rationalise the scenario within a mean field version of the spatial model, that produces a non monotonous flow curve for large restructuring times. This explains the instability of the homogeneous flow below a critical shear rate, that corresponds to the minimum of the curve. Our study therefore strongly supports the idea that the characteristic time scales involved in the local dynamics are at the physical origin of permanent shear bands.}
\vspace{0.5cm}
 \end{@twocolumnfalse}
  ]

\section{Introduction}

\footnotetext{\textit{$^{a}$~LPMCN, Universit\'e de Lyon; UMR 5586 Universit\'e Lyon 1 et CNRS, F-69622 Villeurbanne, France, $^{*}$ kirsten.martens@univ-lyon1.fr.}}
\footnotetext{\textit{$^{b}$~LIPhy, Universit\'e Joseph Fourier Grenoble 1; UMR 5588 et CNRS, F-38402 Saint Martin d'H\`eres, France.}}

Materials that only start to flow beyond a critical yield stress are ubiquitous in nature and serve in a wide range of applications in industry and every day life. Famous examples in soft matter are foams, colloids, dense emulsions, gels, and pastes, but also granular matter and hard amorphous materials like metallic or polymer glasses fall into this category. At the microscopic scale all of these materials have of course very different properties and can barely be described by one single model. However on a more coarse grained level, when for example considering rheological or plasticity features, these materials show very similar properties. 

The steady state rheology in experiments with imposed shear rate $\dot{\gamma}$ yields for example similar flow curves for a variety of complex fluids, that can roughly be categorised in two classes. The first class of yield stress materials will produce rheological curves that can be well fitted by generalised Herschel Bulkley laws, $\sigma=\sigma_y^{(d)}+\alpha\dot{\gamma}^n$. In this description $\sigma$ is the average steady state shear stress and $\sigma_y^{(d)}$ is the dynamical yield stress, corresponding to the low shear rate limit.  The prefactor $\alpha$ and the exponent $n$ are material dependent fitting parameters. Such a law describes a monotonous behaviour of the mean value of the stress as a function of the imposed shear rate. The second class of materials is characterized by flow curves that exhibit a minimum, indicating an instability of homogeneous flow in the region where $d\sigma/d\dot{\gamma}< 0$, and leading to heterogeneous flow in the form of shear bands \cite{Coussot2006}. These two classes have been distinguished in several recent works as ``simple'' fluids without shear banding behaviour and ``thixotropic'' fluids, that phase separate in a solid and a fluid phase \cite{Becu2006, Coussot2007, Bonn2009, Manneville2011}.

It is generally agreed that, as soon as the flow becomes heterogeneous, a description of the rheological behaviour in terms of a flow curve is insufficient.  This has been illustrated  e.g. in the experimental work of Goyon et al \cite{Goyon-Nature, Goyon-SoftMatter}, who showed that a single flow curve was not able to account for the flow profile of a colloidal paste in a microfluidic channel. Such a size dependence of the flow profile hints to the existence of length scales that are not part of the simple rheological description, and several models have recently been put forward to account for the existence of such material specific length scales, either phenomenologically \cite{SGR-shearbanding,ManningLangerCarlson2007} or by an approximate  coarse graining of a mesoscopic description of plastic flow \cite{KEP}. The general structure  couples a local rheological model with a diffusion equation that governs the behaviour of a ''fluidity'' parameter. As a result, coexistence between flowing regions with high fluidity, separated by an interface from regions with low plastic activity, is possible. 

The importance, in terms of strain localisation,  of some kind of local softening mechanism, has been pointed  out  by a number of groups (see e.g. reference \cite{Rodney2011} for a discussion). In  mesoscopic models such as the one we will be discussing below  (see section \ref{sec-model}), softening can arise from an initial ageing process, which results in a distribution of local yield stress that differs from the one in a ''rejuvenated'' system under flow \cite{Vandembroucq2011}. This type of mechanism, that leads to transient (albeit long lived) localisation of strain, was also studied within a ''soft glassy rheology'' context in reference \cite{Moorcroft2011}. The creation of shear bands due to structural relaxation (aging) has also been adressed in a detailed mesoscopic study by Jagla \cite{Jagla2007, Jagla2010}.

A different mechanism, which has been studied at a mean field level by Coussot and Ovarlez \cite{Coussot2010}, involves a local    restructuring time that can be large enough to produce effectively a significant local softening after the system has undergone a local plastic deformation. Similar ideas, introducing directly a local weakening of the yield stress that persists over the duration of avalanches after an initial slip event has occurred, were studied at the mean field level by Dahmen and co-workers \cite{Dahmen2009} and by Mansard {\it et al.} on the basis of a coarse-grained elasto-plastic model \cite{Mansard2011}. It can also be argued that a large restructuring time is, qualitatively, similar to the existence of an efficient relaxation mechanism of the type described in references \cite{Jagla2007, Jagla2010}. In both cases, the local stress will have time to relax significantly before the element is loaded again. The purpose of the study reported in this pa- per is to investigate, beyond the mean field level, the transient softening mechanism proposed by Coussot and Ovarlez \cite{Coussot2010}. 

In the following sections we will show that a long local restructuring time indeed gives rise to the formation of permanent shear bands. We will study  the nature of these bands and show that this heterogeneous flow is different in nature with respect to the formation of dynamical heterogeneities in the simple flow at low shear rates \cite{MBB-PRL2011,BarratLemaitre2011}. In section \ref{sec-model} we introduce the model we consider to describe the local dynamics on a mesoscopic scale. The rheological curves produced by these dynamics are compared to a mean field version of the model in section \ref{sec-rheology}. In this section we motivate the importance of the local restructuring time as model parameter. In section \ref{sec-bandformation} we study the spatial organisation of the plasticity for large restructuring times as a function of the spatial form of the stress propagator, as well as the evolution in time of the shear banding instability. We show that the plasticity organises spontaneously in bands, that persist in the long time limit. The long time characteristics of the dynamics are detailed  in section \ref{sec-longtime}. We show a dependence of the width and the density of the bands on the system size, on the typical restructuring time and on the imposed shear rate. In section \ref{sec-conclusions} we summarize the most important results of our study and give a tentative explanation for the physical origin of the shear band scenario.

\section{Model}
\label{sec-model}
To investigate in detail the influence of the different time-scales in the microscopic dynamics we will study a very simple spatial (as opposed to mean field) model, that accounts for the elasto-plastic behaviour of the medium. The model is essentially the one introduced by Picard and coworkers \cite{Picard2004, Picard2005}. The medium is described by a set of elasto-plastic elements that occupy the nodes of a regular square lattice. To model a yield stress material under steady shear with a fixed strain rate $\dot{\gamma}$, we start from a simple Maxwell-like description of the stress dynamics associated with a given element:
\begin{equation}
\partial_t\sigma(t)=\mu\dot{\gamma}-2\mu\dot{\varepsilon}^\mathrm{pl}(t)
\end{equation}
where $\partial_t$ denotes the partial time derivative, $\sigma(t)$ is the (non tensorial) stress, $\mu$ is the shear modulus, and $\dot{\varepsilon}^\mathrm{pl}(t)$ accounts for the change in the strain due to local yielding.
To take into account the long range effects of the plastic events, first predicted by  Argon \cite{Argon1979} and later evidenced in molecular dynamics simulations \cite{Tanguy2006} and experiments on colloids \cite{Schall2007} we change the above homogeneous equation to a field description with a stress propagator $G(\vec{r}-\vec{r}')$ accounting for the stress redistribution due to a plastic event,
\begin{equation}
 \partial_t\sigma(\vec{r},t)=\mu \dot{\gamma} + 2 \mu \int d\vec{r}' G(\vec{r}-\vec{r}') \dot{\varepsilon}^\mathrm{pl}(\vec{r}',t)
\label{stress-dyn}
\end{equation}
To describe the deformation due to the plastic events, we model a Maxwellian viscoelastic like relaxation of the material in the plastic state, i.e. 
\begin{equation}
\dot{\varepsilon}^\mathrm{pl}(\vec{r},t) = \frac{1}{2\mu\tau}n(\vec{r},t)\sigma(\vec{r},t)\;, 
\end{equation}
where $\tau$ is the typical time for the stress release in the plastic phase and $n(\vec{r},t)$ a local state variable. In the following we refer to $n(\vec{r},t)$ as local activity. We define $n(\vec{r},t)=0$ indicating the absence of a plastic event and $n(\vec{r},t)=1$ if the local region is in the plastic phase.

\begin{figure}[h]
\centering
  \includegraphics[width=1.0\columnwidth,clip]{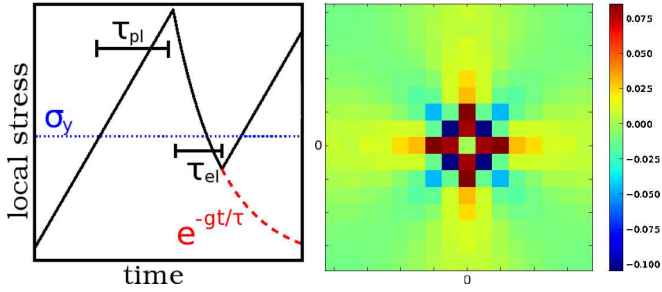}
  \caption{{\it Local stress dynamics:} {\bf Left:} Schematic representation of the local stress of an isolated site as a function of time. When the stress is larger than the threshold $\sigma_y$ the site yields with a probability given by the plastic rate $1/\tau_\mathrm{pl}$. While the site is active the stress drops exponentially during a typical restructuring time $\tau_\mathrm{el}$ until it becomes elastic again. The site is supposed to be isolated, to simplify the picture. In the completely connected model described above we would expect stress fluctuations due to the mechanical noise coming from the surrounding sites. {\bf Right:} Stress change due to a single plastic event at the origin. The stress change at the origin is artificially set to zero to account for the visibility of the nearest neighbour changes. In reality the change at the origin is strongly negative.}
  \label{fig-stressdyn}
\end{figure}

To incorporate the idea of a yield stress, we define the following stochastic dynamics for the state variable $n(\vec{r},t)$ (see left panel of Fig.~\ref{fig-stressdyn}). If the local stress exceeds a threshold value $\sigma(\vec{r},t)>\sigma_y$ there is a finite probability to yield locally. The yielding rate is given by $1/\tau_\mathrm{pl}$. Once yielded ($n(\vec{r},t)=1$) stress is redistributed using an appropriate stress propagator and locally the stress is relaxing during a typical restructuring time $\tau_\mathrm{el}$. The rate to re-establish the elastic state ($n(\vec{r},t)=0$) is given by $1/\tau_\mathrm{el}$. Note that in former studies of this model both rates have been set equal to the elementary relaxation time \cite{Picard2005}. In this work we focus on the dependence of the dynamics on the ratio of these time scales.

Approximating the rearrangements corresponding to the local plastic events by a force quadrupole (Eshelby problem), one expects a four-fold quadrupolar symmetry for the inhomogeneous part of the stress propagator \cite{Eshelby} (see right panel of Fig.~\ref{fig-stressdyn}). The form of this propagator for an infinite two dimensional medium reads
\begin{equation}
G^{(\infty)}(r,\theta)=\frac{1}{\pi r^2}\cos(4\theta)\;.
\label{long-range-prop}
\end{equation}

In the following we will restrict our interest to athermal systems in a planar geometry. Further, to simplify the problem we consider an incompressible medium and assume that the microscopic geometry of the plastic events is the same as the one of the macroscopic shear. This simplification permits to use the above introduced scalar description of the stress. Note that convection effects on the stress and the activity have been neglected to keep the model as simple as possible. The lack of convection leads basically to an additional x-y symmetry for the spatial arrangements of the plastic events. For the shear bands this means that there is no preferred direction of the flow, such that they will be oriented with equal probability in the $x$ or $y$ direction.

To simplify, we rewrite Eq.~(\ref{stress-dyn}) in dimensionless quantities, where the time $t$ is measured in units of the elementary relaxation time $\tau$, the stress $\sigma$ in units of the local yield stress $\sigma_y$ and the strain rate $\dot{\gamma}$ in units of the critical  value $\dot{\gamma}_c=\sigma_y/\tau\mu$.  $\dot{\gamma}_c$ indicates the change from a Newtonian to a non Newtonian flow behaviour. To concentrate on the bulk dynamics we implement periodic boundary conditions and we integrate Eq.~\ref{stress-dyn} numerically following the approach of Picard et al.\cite{Picard2004, Picard2005}. We encode the local stress values $\sigma_{ij}$ ($i$ and $j$ being the discretized space coordinates) on a square lattice of size $N=L^2$, given in units of $a^2$, the typical size of a plastic event. Further we calculate the convolution in the second term of Eq.~\ref{stress-dyn} using fast Fourier transform methods. We approximate the finite size propagator corresponding to Eq.~\ref{long-range-prop} in Fourier space as
\begin{equation}
 \hat{G}(q_i,q_j)=-4 \frac{q_i^2 q_j^2}{(q_i^2+q_j^2)^2}
\end{equation}
with wave vector $\vec{q} = (q_i, q_j) = \frac{2\pi}{L} (i,j)$. The homogeneous part of the propagator is implemented as $ \hat{G}(0,0) = -1$. The integration time step is taken to be small with respect to all other time scales, $\Delta t/\tau=10^{-2}$.

The numerical implementation of the stochastic yielding dynamics is straightforward. To be precise, we summarize the above described dynamics for the activity field $n_{ij}$ in the following expressions
\begin{eqnarray}
\label{stoch-eq}
P_{01}(n_{ij}(t+dt)=1 | n_{ij}(t)=0; \sigma_{ij}>\sigma_y) &=& \frac{dt}{\tau_\mathrm{pl}} \\
P_{10}(n_{ij}(t+dt)=0 | n_{ij}(t)=1) &=& \frac{dt}{\tau_\mathrm{el}} \;,
\end{eqnarray}
where $P_{01}(n(t+dt)=1|n(t)=0; \sigma>\sigma_y)$ denotes the transition probability of going from state $n=0$ to state $n=1$ in a small time interval $dt$, given that the local stress value exceeds the yield stress. $P_{10}(n(t+dt)=0|n(t)=1)$ denotes the corresponding reverse transition probability.

After the rescaling we obtain a minimal model describing the dynamics of two fields, namely the stress field $\sigma_{ij}$ and the activity field $n_{ij}$ with only a few parameters, that all have a physical motivation. The system size $N$ and the shear rate $\dot{\gamma}$ are macroscopic controllable quantities. Further there are the different time-scales involved in the microscopic dynamics, $\tau_\mathrm{pl}$ and $\tau_\mathrm{el}$, describing the typical time to yield and the typical restructuring time respectively. The yielding time $\tau_\mathrm{pl}$ accounts for the disorder in the system. Defining a probability to yield above the globally defined yield stress is equivalent to introduce random barriers in the system, that change once a site has yielded. This is  similar to the definition of the soft glassy rheology model \cite{SGR97}, but with a strain rate dependent probability distribution for the barrier heights. The simulations presented in this paper have been performed for a value of the typical time to yield equal to the elementary relaxation time ($\tau_\mathrm{pl}=1$).

The most interesting control parameter in this study is however the restructuring time $\tau_\mathrm{el}$. This parameter determines how long the system is locally in a plastic state after a yielding event. The larger $\tau_\mathrm{el}$ the longer we relax locally the stress during a plastic event and redistribute it over the system. In a more physical interpretation we expect this parameter to correspond to the time the material needs to regain its original structure, after links are broken trough the local rearrangements in the plastic event. This interpretation is similar to what has been recently proposed by Coussot et al.~\cite{Coussot2010} within a mean field model for shear banding behaviour. In the following sections we will show that this parameter plays an essential role in the formation of a shear band structure. 

\section{Rheology}
\label{sec-rheology}
Before going into the study of the spatial organisation of the plasticity within our model, we shall study some macroscopic features, such as the flow rheology. To do so we derived an analytically solvable mean field description for our model, in which we describe the effective dynamics on a single site. We rewrite the stress dynamics of the full model as
\begin{equation}
 \partial_t \sigma_{ij}(t) = \dot{\gamma}^\mathrm{eff}_{ij}(t) + G(0,0) n_{ij}(t) \sigma_{ij}(t)\;,
\label{effective-dynamics}
\end{equation}
where  $G(0,0)<0$ is the value of the stress propagator at the origin;
$\sigma_{ij}(t)$ denotes the shear stress at point $\{i,j\}$ and 
\begin{equation}
  \dot{\gamma}^\mathrm{eff}_{ij} =  \dot{\gamma} + \sum_{i'j' (\neq ij)} G(i-i',j-j') n_{i'j'} \sigma_{i'j'}\;.
\label{eff-gamma_p-def}
\end{equation}

If we assume $\dot{\gamma}^\mathrm{eff}_{ij}$ to be homogeneous in space and if we neglect any possible fluctuations around its mean value $\dot{\Gamma}$ we obtain the following effective dynamics for the stress evolution on one site
\begin{equation}
\label{mf}
 \partial_t \sigma(t)= \dot{\Gamma} -g n(t) \sigma(t)
\end{equation}
with $g=-G(0,0)>0$ the absolute value of the stress propagator at the origin. This value is system size dependent in the fully spatial model, but approximately constant in the limit of large systems. In the range studied within this paper g is large with respect to the shear rates of interest ($g\approx 0.57$). The activity $n(t)$  obeys the same local stochastic rules as in the full spatial dynamics, given in equation (\ref{stoch-eq}). The advantage of this mean field model is that we can extract analytically the expression for the flow curve, that is the average steady state stress as a function of the effective shear rate $\dot{\Gamma}$. 

To derive an approximate solution for the mean field equation (\ref{mf}), we assume $\dot{\Gamma}<g$ and $\tau_\mathrm{el}$ sufficiently large to decorrelate the typical stress values for the change of activity $n$. We have the following two equations that describe the average dynamics in the inactive ($\overline{n}=0$) and in the active ($\overline{n}=1$) part of the dynamics:
\begin{eqnarray}
 \overline{n}=0 : \quad \overline{\sigma}(t)&=&\overline{\sigma}_{-} +\dot{\Gamma} t\\
 \overline{n}=1 : \quad \overline{\sigma}(t)&=&\frac{\dot{\Gamma}}{g} + A(\dot{\Gamma}, \tau_\mathrm{pl}) e^{-g t}
\end{eqnarray}
where the bar indicates an ensemble average. Further $\overline{\sigma}_{-}$ is given by the typical stress value at which the site becomes elastic 
\begin{equation}
 \overline{\sigma}_{-}=\frac{\dot{\Gamma}}{g}+\frac{A(\dot{\Gamma},\tau_\mathrm{pl})}{1+g\tau_\mathrm{el}}
\end{equation}
and the prefactor $A(\dot{\Gamma}, \tau_\mathrm{pl})= \overline{\sigma}_{+}-\dot{\Gamma}/g$ is fixed by
\begin{equation}
 \overline{\sigma}_{+}=1+\dot{\Gamma}\tau_\mathrm{pl}\;,
\end{equation}
the typical stress value at which the site yields. 

To calculate the time averaged stress $\langle \sigma \rangle_t$, we now aim for the analytical expressions for the typical times spent in the two different activity states and the associated average stress values
\begin{eqnarray}
 \langle\sigma\rangle_t = \frac{\tau_\mathrm{in} \langle \sigma_\mathrm{in}\rangle_t+\tau_\mathrm{act} \langle\sigma_\mathrm{act}\rangle_t}{\tau_\mathrm{in}+\tau_\mathrm{act}}\;.
\end{eqnarray}

The time average stress value in the inactive phase is given by 
\begin{eqnarray}
 \langle\sigma_\mathrm{in}\rangle_t &=& \frac{1}{\tau_\mathrm{in}}\int d\sigma_{+} P_+(\sigma_{+})\int d\sigma_{-} P_-(\sigma_{-})\times\nonumber\\
&& \qquad\left(\frac{\sigma_{+}+\sigma_{-}}{2}\right)\left(\frac{\sigma_{+}-\sigma_{-}}{\dot{\Gamma}}\right) \;,
\end{eqnarray}
where $P_+$ and  $P_-$ denote the probability distributions of $\sigma_{-}$ and $\sigma_{+}$ respectively. 
The typical corresponding time, spent in the inactive phase simply yields $\tau_\mathrm{in}=(\overline{\sigma}_{+}-\overline{\sigma}_{-})/\dot{\Gamma} = \tau_\mathrm{pl}+(1-\overline{\sigma}_{-})/\dot{\Gamma}$. 
Therefore we obtain for the average stress in the inactive state
\begin{equation}
 \langle\sigma_\mathrm{in}\rangle_t = \frac{\overline{\sigma_{+}^2}-\overline{\sigma_{-}^2}}{2\left(\overline{\sigma}_{+}-\overline{\sigma}_{-}\right)}\;,
\end{equation}
with
\begin{eqnarray}
 \overline{\sigma_{+}^2}&=&1+2\dot{\Gamma}\tau_\mathrm{pl} \overline{\sigma}_+\\
 \overline{\sigma_{-}^2}&=&\left(\frac{\dot{\Gamma}}{g}\right)^2+2\frac{\dot{\Gamma}}{g} \frac{A(\dot{\Gamma},\tau_\mathrm{pl})}{1+g\tau_\mathrm{el}}\\
&&+\frac{\overline{\sigma_+^2}-2 (\dot{\Gamma}/g)\overline{\sigma}_++(\dot{\Gamma}/g)^2}{1+2g\tau_\mathrm{el}}\nonumber\;.
\end{eqnarray}

The last missing information is the average stress value in the active state, $\sigma_\mathrm{act}$ which can be calculated by
\begin{eqnarray}
\label{simga_act}
 \langle\sigma_\mathrm{act}\rangle_t&=& \frac{1}{\tau_\mathrm{el}^2}\int_0^{\infty} d\tau e^{\tau/\tau_\mathrm{el}} \int_{0}^{\tau} dt'\left(\frac{\dot{\Gamma}}{g}+Ae^{-gt'}\right)\nonumber\\
&=&\frac{\dot{\Gamma}}{g}+\frac{A(\dot{\Gamma}, \tau_\mathrm{pl})}{1+g\tau_\mathrm{el}}\;.
\end{eqnarray}
This means that the average stress in the active state is  equal to the typical stress value to become elastic $\langle\sigma_\mathrm{act}\rangle_t=\overline{\sigma}_-$. The typical time spent in the active state is simply given by $\tau_\mathrm{act}=\tau_\mathrm{el}$. 

\begin{figure}[h]
\centering
\includegraphics[width=\columnwidth,clip]{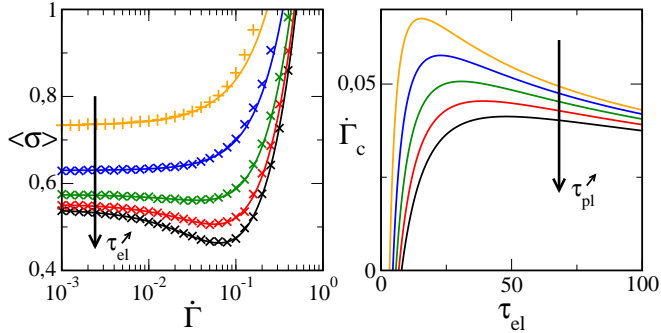}
\caption{{\it Rheological curves:}  {\bf Left:} Shown are the numerically measured values of the steady state stress $\langle \sigma \rangle$ of the mean field model as a function of strain rate $\dot{\Gamma}$ for different values of the restructuring time ($\tau_\mathrm{el} = 1, 2.5, 5, 7.5, 10$) for $\tau_\mathrm{pl}=1$. This is compared to the analytically derived expressions for the stress dependence on the strain rate for the same restructuring times (solid lines). {\bf Right:} Strain rate $\dot{\Gamma}_c$ corresponding to the minima of the mean field flow curves in the left panel as a function of restructuring time $\tau_\mathrm{el}$ for different values of the yielding rate ($\tau_\mathrm{pl}=1.0, 2.0, 3.0, 4.0, 5.0$).}
  \label{fig-flowcurve}
\end{figure}

Using the analytical expressions in the former paragraph we can express the steady state average stress as a function of strain rate.
In  Figure \ref{fig-flowcurve} we show the steady state solution of these mean field equations for various values of the restructuring time $\tau_\mathrm{el}$. We find that above a critical value of this restructuring time $\tau_\mathrm{el}^c$ the mean field solution exhibits a minimum. This means for a given value of the average stress $\langle \sigma\rangle$ we find more than one value of the effective shear rate, which may give rise to phase coexistence in the spatial model. From the above equations the critical value for the restructuring time can be derived through $\lim_{\dot{\Gamma}\to 0}d\langle\sigma\rangle/d\dot{\Gamma}=0$, yielding
\begin{equation}
\label{tau-el-crit}
 \tau_\mathrm{el}^c=\frac{1+g\tau_\mathrm{pl}+\sqrt{1+6g\tau_\mathrm{pl}+g^2\tau_\mathrm{pl}^2}}{2g}\;.
\end{equation}
For a typical yielding rate $\tau_\mathrm{pl}^{-1}=1$ and an absolute value of the propagator at the origin of $g=0.57$ this yields a critical restructuring time $\tau_\mathrm{el}^c\approx 3.3$. For restructuring times above this critical value we obtain a minimum in the flow curve within the presented mean field description. There is a critical strain rate $\dot{\Gamma}_c$ below which the rheological curve exhibits a negative slope. The value of this critical shear rate will both depend on the restructuring time $\tau_\mathrm{el}$ and on the yielding time $\tau_\mathrm{pl}$ (see right panel of Figure \ref{fig-flowcurve}).

Another quantity of interest, that can be easily derived in this mean field approximation, is the average steady state activity $\langle n\rangle_t$ as a function of effective shear rate $\dot{\Gamma}$, given by $\langle n\rangle_t=\tau_\mathrm{act}/(\tau_\mathrm{act}+\tau_\mathrm{in})$. This yields
\begin{equation}
\label{eq-activity}
 \langle n\rangle_t=\frac{1+g\tau_\mathrm{el}}{g(\tau_\mathrm{el}+\tau_\mathrm{pl}+1/\dot{\Gamma})}\;.
\end{equation}
An alternative way to derive the average activity is to consider the steady state state expression $\langle d_t \sigma \rangle_t =0$ and thus $\langle n\sigma \rangle_t=\dot{\Gamma}/g$, which can be rewritten as $\langle n \rangle_t \sigma_\mathrm{act}= \dot{\Gamma}/g$, recovering the same expression (\ref{eq-activity}).

The main difference between our mean field model and the former introduced model of Coussot et al.~is the existence of a third time scale, which is the typical rate to yield $\tau_{pl}$. In the model of Coussot et al, the flow behaviour turns from that of a simple yield stress fluid to a shear-banding material when the ratio of a characteristic relaxation time of the system to a restructuring time becomes smaller than one \cite{Coussot2010}. Note that in the limit of $\tau_\mathrm{pl}=0$ and a value of $g=1$, we recapture this result within our mean field model (see Eq.~\ref{tau-el-crit}). The flow curve however is different from the simple mean field result in the work of Coussot and coworkers, due to the more complex relaxation dynamics (exponential decay of the local stress).

\section{Band formation}
\label{sec-bandformation}
\begin{figure}[h]
\centering
  \includegraphics[width=1.0\columnwidth,clip]{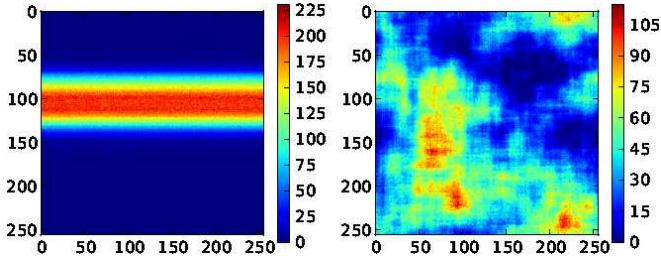}
  \caption{{\it Cumulated activity:} Local number of plastic events in a given strain interval $\Delta\gamma=100$ in a system of size $N=2^{16}$ with a restructuring time $\tau_\mathrm{el}=10.0$. {\it Left:} stress is redistributed using the standard long range stress propagator, given in Eq.~(\ref{long-range-prop}). {\it Right:} stress dynamics are governed by an alternative short range stress propagator (see text). }
  \label{fig-accumulated}
\end{figure}

From the above observed minimum in the rheological curve we expect that the activity localises in some specific regions. But are the plastic events really concentrated in shear bands? One way to get an idea about the spatial organisation of the plastic events is to calculate numerically the accumulated activity map, that is the local number of plastic events in a given strain window. It turns out that the specific form of the stress propagator plays an important role for the spatial distribution of the plastic events. When the   
original long range propagator (given in Eq.\ref{long-range-prop}) is used to redistribute the stresses after a plastic event occurred, we indeed obtain a shear band structure (see right panel of Fig.\ref{fig-accumulated}). However if we use a simplified short range propagator with the same homogeneous part but an inhomogeneous stress redistribution, given by the following table:
\begin{center}
\begin{tabular}{c|c|c}
-1/4 & 1/2 & -1/4 \\
\hline
1/2 & -1 & 1/2 \\
\hline
 -1/4 & 1/2 & -1/4 
\end{tabular} 
\end{center}
we obtain active regions of typical size, but of a shape that is very different from the band geometry (see right panel of Fig.~\ref{fig-accumulated}). This means that the long range character of the propagator enhances the shear banding effect.  This can be understood from the fact that, with a short range propagator, a plastically active  region of arbitrary shape can coexist with an  inactive region, with a smooth interface separating the two regions. For the long range elastic propagator, on the other hand,  a uniform line of slip events is the only type of plastically active region that does not give rise to any deformation in the surrounding medium.

\begin{figure}[h]
\centering
  \includegraphics[width=0.8\columnwidth]{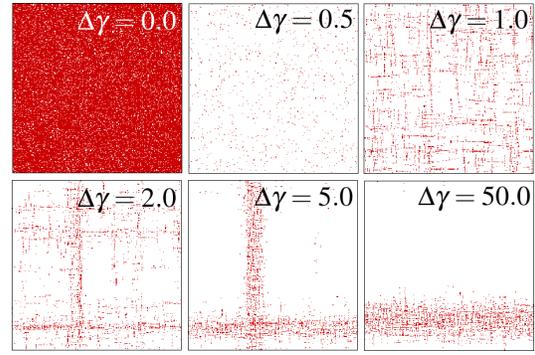}
  \caption{{\it Quench dynamics:} Instantaneous activity maps $n_{ij}$, with $i,j \in [0,L]$ at different strain intervals $\Delta\gamma$ after a rapid quench of the shear rate from $\dot{\gamma}=1.0$ to $\dot{\gamma}=10^{-2}$ for a system size $N=2^{16}$ and a restructuring time of $\tau_\mathrm{el}=10$. Active regions are indicated in red ($n_{ij}=1$), passive regions in white  ($n_{ij}=0$).}
  \label{fig-quench}
\end{figure}

To test whether these bands are only a transient phenomena leading to a homogeneous state in the long time limit, we performed the following simulation. At the beginning we apply a high strain rate $\dot{\gamma}=1$ where the dynamics is homogeneous, that is the probability of producing a plastic event in the stationary state is uniform in space. Then we perform a rapid quench of the shear rate into the unstable regime with $\dot{\gamma}=0.01$. This allows to start from a homogeneous condition. Moreover due to the periodic boundary conditions and homogeneous applied shear rate, there is no way to break the homogeneity except for the appearance of instabilities due to collective effects in the stress dynamics. And indeed we observed the appearance of thin and short shear bands and an ensuing coarsening that lead to longer and wider bands, eventually reaching the system size (see Fig.~\ref{fig-quench}). In the long time limit one system size spanning band is formed, that has a characteristic width and density. This band can drift in perpendicular direction of its orientation.

Note that for very small system sizes, strong fluctuations can lead to a renucleation of the band in another region. But even in this regime the structure factor remains constant and we obtain a permanent instability. For very large system sizes we observe a transient regime in which several bands can coexist during a finite time. However, for the parameter range studied here we always obtain a steady state (with a time independent static structure factor) producing one single band that only moves perpendicular to its orientation. The system size is chosen sufficiently large to ensure that fluctuations will not lead to a renuclation of the band elsewhere in the system.

\section{Long time characteristics of the shear bands}
\label{sec-longtime}
In this section we will discuss the long time characteristics of the formed bands. As we showed in the former section the spatial organisation of the activity in the long time limit corresponds to one single band of a given width, density and interface width. These  characteristics of the band will depend on the model parameters, that is system size $N$, restructuring time $\tau_\mathrm{el}$ and strain rate $\dot{\gamma}$.

\begin{figure}[h]
\centering
\includegraphics[width=\columnwidth,clip]{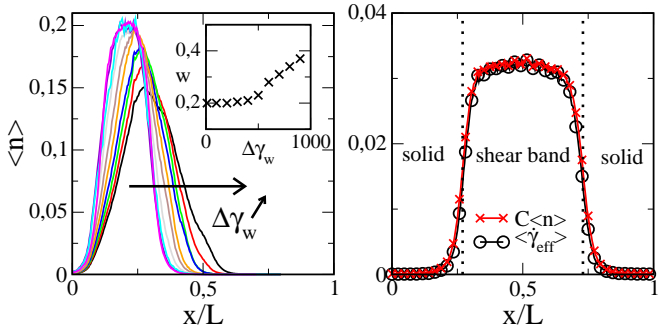}
  \caption{{\it Band characteristics} {\bf Left:} Average activity $\langle n \rangle$ as a function of space (coordinate perpendicular to the band orientation). The average is taken over the second space coordinate and over a given strain window $\Delta \gamma_w$ (10, 100, 200, 300, 400, 500, 600, 700, 800, 900). The inset shows the rescaled position of the band $x_p/L$ as a function of strain from band profiles obtained through averaging over a strain window of $\Delta \gamma_w=10$. {\bf Right:} Average local effective shear rate $\langle\dot{\gamma}_\mathrm{eff}\rangle$and rescaled average local activity $\langle n\rangle/C$ as a function of space for restructuring time $\tau_\mathrm{el}=10$, shear rate $\dot{\gamma}=10^{-1.6}$ and system size $N=2^{16}$. The numerically measured value of the rescaling factor is $C=0.155$ (see text).}
  \label{fig-band-measure}
\end{figure}

But before studying in detail these dependences we need to give a definition of what we want to call a permanent shear band. To obtain a characteristic band profile we need to average over a certain strain window $\Delta \gamma_w$. This strain window must be large enough to capture the dynamically built shear band, but small with respect to the typical strain that causes the band to change its position. In the left panel of Fig.~\ref{fig-band-measure} we show the band profile as a function of different strain windows, used for the averaging. Indeed, we find that it is possible to obtain a band profile that is well defined for a range of averaging strain windows, as long as these are small with respect to the strain window that leads to a significant band displacement (perpendicular to the band orientation). 

To obtain the long time characteristics of a band we use a gliding window average in the steady state, where we translate the band, obtained in each window, to the same position (centre of the system) for the averaging. In the right panel of Fig.~\ref{fig-band-measure} we show such a band profile for given model parameters that yield inhomogeneous flow. Indeed we find a region with a strong activity that we call shear band and regions with almost no plastic events, that we will call solid phase in the following.

But how does the profile of the average activity $\langle n\rangle$ compare to the profile of the average effective shear rate? 
To obtain this relation we study the steady state dynamics where $\partial_t \langle \sigma \rangle=0$, which yields, using relation (\ref{prop-sum})
\begin{eqnarray}
 \dot{\gamma}+\frac{1}{L^2}\langle \sum_{ii'jj'} G({i-i',j-j'} n_{i'j'} \sigma_{i'j'} \rangle_t&=&0\nonumber\\
 \langle n\sigma \rangle &=& \dot{\gamma}\;.
\end{eqnarray}
This reflects the fact that the space averaged change of plastic strain equals the globally applied shear rate, leading to a non-equilibrium steady state.

\begin{figure}[h]
\centering
  \includegraphics[width=1.0\columnwidth,clip]{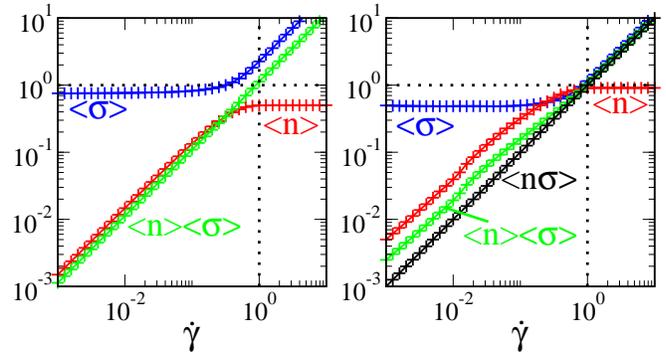}
  \caption{{\it Flow and activity curves:} {\bf Left:} Shown are the average stress $\langle\sigma\rangle$, the average activity $\langle n\rangle$ and the product of both $\langle n\rangle\langle\sigma\rangle$ as a function of strain rate $\dot{\gamma}$, for a system size $N=2^{16}$ and a value of the restructuring time $\tau_\mathrm{el}=1$ (homogeneous flow).
{\bf Right:} Shown are the average stress $\langle\sigma\rangle$, the average activity $\langle n\rangle$, the product of both $\langle n\rangle\langle\sigma\rangle$ and the average stress weighted by the activity of the corresponding site $\langle n\sigma\rangle$ as a function of strain rate $\dot{\gamma}$, for a system size $N=2^{16}$ and a value of the restructuring time $\tau_\mathrm{el}=10$ (inhomogeneous flow).}
  \label{fig-correlations}
\end{figure}

Further it turns out that for homogeneous flow, $\tau_\mathrm{el}<\tau_\mathrm{el}^c$ 
activity and stress are uncorrelated, such that $\langle n \rangle \langle \sigma \rangle = \dot{\gamma}$ (see left panel of Figure \ref{fig-correlations}. This is in agreement with the KEP model assumption \cite{KEP}, if we interpret the local activity as fluidity. However in the inhomogeneous flow regime there are strong negative correlations between activity and stress (see right panel of Figure \ref{fig-correlations}). This is intuitive since an active site that remains active for a long time will lead to a strong local stress relaxation.

To get an estimate of the relation between the effective shear rate and the activity, we study equation (\ref{effective-dynamics}) in the steady state regime where we find
\begin{eqnarray}
 \langle \dot{\gamma}_{ij}^\mathrm{eff}\rangle_t&=&g\langle n_{ij}\sigma_{ij}\rangle_t\nonumber\\
&\approx&g\langle n_{ij}\rangle_t \sigma_\mathrm{act}\;,
\end{eqnarray}
where $\sigma_\mathrm{act}$ indicates the time averaged stress value on an active site. 
We find that the local average effective shear rate is linear in the local average activity, $\langle \dot{\gamma}_{ij}^\mathrm{eff}\rangle_t\approx C \langle n_{ij}\rangle_t$, with $C=g \sigma_\mathrm{act}$.
As mentioned earlier the value of $g$ depends on the systems size. For example for the parameters $N=2^{16}$, $\dot{\gamma}=10^{-1.6}$ and $\tau_\mathrm{el}=10$ the proportionality constant yields $C\approx0.155$, see Fig.\ref{fig-band-measure}.

It  immediately follows that in the regions where $\langle n_{ij}\rangle \approx0$ we also obtain a vanishing small effective local shear rate. We will refer to these regions as solid regions. On the contrary in the regime with high activity we obtain a finite effective shear rate corresponding to a shear band (see right panel of Fig \ref{fig-band-measure}). In the following we will use the activity rather than the measurement of the effective shear rate for the definition of the shear bands.

\begin{figure}[h]
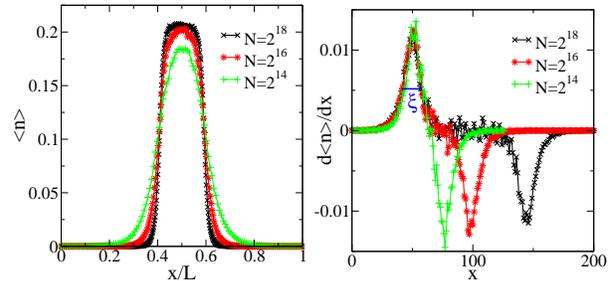

\centering
  \includegraphics[width=0.45\columnwidth,clip]{figure7a.eps}
  \includegraphics[width=0.45\columnwidth,clip]{figure7b.eps}
  \caption{{\it System size dependence of the band profile:} {\bf Left:} System size dependence of the band profile. Shown is the average activity $\langle n \rangle$ in the steady state as a function of space (coordinate always perpendicular to the band orientation) rescaled with system length $L$ for different values of the system size $N=L^2$. The averaging strain window is $\Delta\gamma_w=10$ and we average over a total strain of $\Delta\gamma=1000$. {\bf Right:} Space-derivative of the average activity $\langle n \rangle$ as a function of space for different values of the system size. The position of the band is shifted, such that the beginning of the band is in the same position. The length $\xi$ is a measure of the width of the interface.}
  \label{fig-sizedep}
\end{figure}

Let us now study in detail the long time characteristics of the band as a function of system size, shear rate and restructuring time.
In the left panel of Fig.~\ref{fig-sizedep} we display this band profile as a function of system size for $\tau_\mathrm{el}=10$ and $\dot{\gamma}=10^{-2}$. It turns out that the width of the band is linear in system size and the width of its interface $\xi$  has no significant dependence of the size of the system (see right panel of Fig.~\ref{fig-sizedep}).

\begin{figure}[h!]
\centering
\includegraphics[width=\columnwidth,clip]{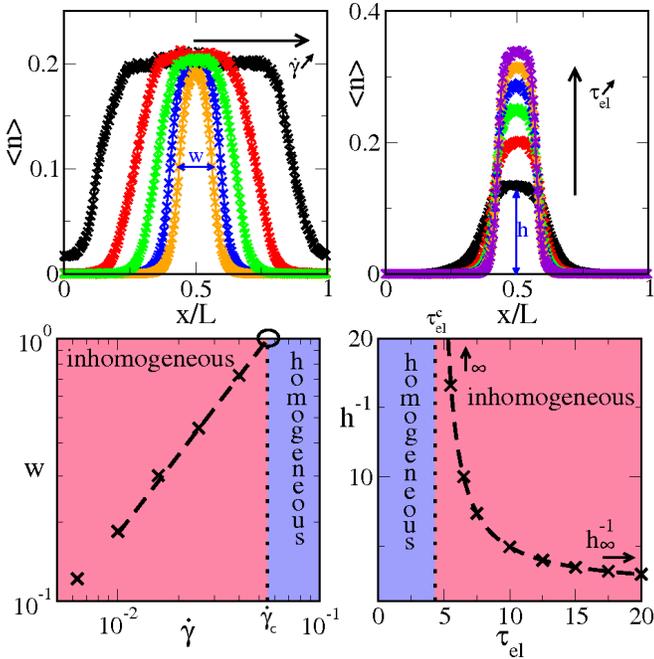}
  \caption{{\it The flow transition:} {\bf Upper left:} Band profile of the single band forming in the long time limit as a function of the shear rate $\dot{\gamma}$ ($ 10^{-2.2}, 10^{-2.0}, 10^{-1.8}, 10^{-1.6}, 10^{-1.4}$) for a restructuring time $\tau_\mathrm{el}=10$. {\bf Upper right:} Band profile  as a function of the restructuring time $\tau_\mathrm{el}$ (5.5, 6.5, 7.5, 10, 12.5, 15, 17.5, 20) for a shear rate $\dot{\gamma}=10^{-2}$. {\bf Lower left:} Width $w$ (rescaled with system length) as a function of $\dot{\gamma}$ for $\tau_\mathrm{el}=10$. The dashed line is a linear fit of the large shear rate region and the black circle indicates the point where we expect the transition to homogeneous flow (the width of the band reaches the system size at $\dot{\gamma}_c \approx 5.5\times 10^{-2}$). The dotted line separates the region of homogeneous from the region of inhomogeneous flow. {\bf Lower right:} Inverse peak activity  $h^{-1}$ of the band profile as a function of the restructuring time $\tau_\mathrm{el}$ for $\dot{\gamma}=10^{-2}$. The dashed line is a fit (see text) and the dotted line indicates the fitted critical restructuring time for the transition from homogeneous to inhomogeneous flow behaviour at $\tau_\mathrm{el}^c\approx 4.3$ (see text).}
  \label{fig-characteristics}
\end{figure}

As for what is the dependence of the band characteristics on the shear rate, we find that the maximum (or peak) activity of the band does not depend on $\dot{\gamma}$. The peak activity $h$ is here defined as the difference between the minimal and the maximal value of the average activity.
The average effective shear rate in the shear band $\dot{\gamma}_\mathrm{eff}^{(b)}$ takes the value where the average stress in the corresponding effective flow curve is minimal (see Figure \ref{fig-flowcurve}). This fixes the average stress in the solid phase as well. It is only the relative width of the band $w$ that changes with $\dot{\gamma}$. For large $\dot{\gamma}$ the dependence is linear. This allows for an extrapolation, yielding the critical shear rate $\dot{\gamma}_c \approx 5.5\times 10^{-2}$ at which the width of the band reaches the system size. This is the point of the crossover from a heterogeneous flow to a homogeneous flow without any shear bands (see left panels of Figure \ref{fig-characteristics}.

When changing the value of the restructuring time, we observe a different phenomenon. For large $\tau_\mathrm{el}$ the width of the band is fixed, but the peak activity $h$ of the band grows with increasing $\tau_\mathrm{el}$. On the other hand below a critical value of the restructuring time $\tau_\mathrm{el}^c$ the band vanishes and the flow becomes homogeneous. To estimate this crossover time, we use that in the steady state $\langle n\rangle \sigma_\mathrm{act}$ equals $\dot{\gamma}$. In the parameter range where the width is fixed the peak activity  must compensate the decreasing value for the average stress on the active sites $\sigma_\mathrm{act}$ due to the long time of local relaxation. A simple fitting function for the inverse peak activity is 
\begin{equation}
 h^{-1}=\frac{a(\dot{\gamma})}{\tau_\mathrm{el}-\tau_\mathrm{el}^c} +h^{-1}_\infty
\end{equation}
with the fitting parameters $h^{-1}_\infty>1$, being the peak activity at large restructuring times, $a(\dot{\gamma})$, a shear rate dependent prefactor and $\tau_\mathrm{el}^c$, the critical restructuring time for which the system becomes homogeneous. With this fitting procedure we find a value of the critical restructuring time of $\tau_\mathrm{el}^c\approx 4.3$, which is in fair agreement with the findings within the mean field model, see Eq.~(\ref{tau-el-crit}). Note, that for very large $\tau_\mathrm{el}$ the time to form one single band grows strongly and becomes inaccessible. Instead of one large band we observe several smaller coexisting bands. Thus, the above relations are only valid in a finite range of values for $\tau_\mathrm{el}$.

\section{Discussion and conclusions}
\label{sec-conclusions}

We performed the first detailed spatial analysis of the formation of shear bands within a minimalistic model, that bridges the microscopic dynamics to the macroscopic flow behaviour. The model for the yield stress flow invokes only few physically relevant parameters from the microscopic and from the macroscopic scale. 

We find that a microscopic time-scale, that is linked to the restructuring time of the material, controls the crossover from homogeneous to heterogeneous flow, similar to what has recently been predicted within a mean field study \cite{Coussot2010}. If the time needed after a local plastic event, to regain the original structure is large in comparison with the typical duration of local stress release, the system undergoes a phase separation below a critical shear rate. The separation gives rise to a strongly sheared phase of band geometry within a solid non flowing environment. We show that this band geometry is closely related to long range character of the stress redistribution due to a local plastic event.

After a quench from a large to a small shear rate we observe coarsening behaviour for the activity in the system, that yields in the long time limit a single strongly active band. The band characteristics are well defined and depend on the system size, the value of the shear rate and the typical local restructuring time. We study these dependences in detail and show how to get a prediction for the relevant critical values at which the crossover from simple to shear band flow occurs. 

These findings are compared to a mean-field version of the model that incorporates the influence of all other surrounding sites into an effective shear rate. There are two important questions to answer with respect to the spatial model to be able to make predictions about the phase coexistence. Let us assume that our dynamics produce two different phases in form of a shear band and inactive {\it solid} regions. Then we have to show that the average stress will be the same in the two phases. And indeed the stress dynamics are defined such that if we start from an homogeneous initial condition for the stress field, the dynamics will assure that the average stress in a line or in a column is equal to the global average stress. This is due to the specific form of the stress propagator, that is defined such that
\begin{equation}
 \sum_i G(i,j) = \sum_j G(i,j) = -\frac{1}{L} \;.
\label{prop-sum}
\end{equation}
The change of the average stress on one line (horizontal or vertical) during one time step $\Delta t$ turns out to be independent of the position of the line
\begin{eqnarray}
 \frac{1}{L}\sum_{i} \Delta \sigma_{ij} &=&\left( \dot{\gamma} + \frac{1}{L}\sum_{ii'j'} G(i-i',j-j')n_{i'j'}\sigma_{i'j'}\right) \Delta t\nonumber\\
\langle\Delta\sigma_{ij}\rangle_i&=&\left(\dot{\gamma}-\langle n_{ij}\sigma_{ij}\rangle_{ij}\right)\Delta t\; ,
\end{eqnarray}
which is independent of $j$. The same argument is valid for summation over space coordinate $i$. If we start from a homogeneous condition for the stress values (in the simulations $\sigma_{ij}=0$, $\forall ij$) the average stress on a line will always be equal to the global average of the stress in the system. Thus, by construction the average stress in a shear band will equal the average stress in the inactive phase. The average stress can therefore be interpreted as intensive thermodynamic parameter that governs the phase coexistence.

\begin{figure}[h!]
\centering
\includegraphics[width=\columnwidth,clip]{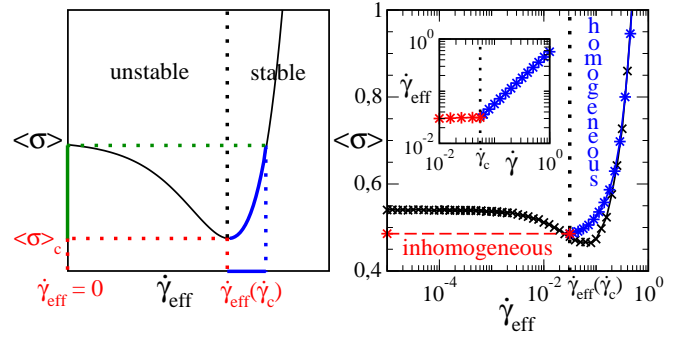}
\caption{{\it Phase separation scenario:} {\bf Left:} Sketch of a typical mean field flow curve that leads to a phase separation. The green line indicates all possible values for the stress and the blue line all possible values for the high shear rate phase.  In red are indicated the dynamically chosen values for the phase coexistence, $\dot{\gamma_\mathrm{eff}}=0$ and $\dot{\gamma_\mathrm{eff}}(\dot{\gamma}_c$) at equal stress $\langle\sigma\rangle_c$, that correspond to the minimum of the flow curve. {\bf Right:} Average steady state stress value $\langle\sigma\rangle_{ss}$ as a function of the effective strain rate $\dot{\gamma}_\mathrm{eff}$ for a value of the restructuring time $\tau_\mathrm{el}=10$ in the complete model with $N=2^{16}$ ($\ast$) in comparison with the flow curve of the mean field version of the model ($\times$). The vertical dotted line indicates the critical value of the effective shear rate $\dot{\gamma}_\mathrm{eff}(\dot{\gamma}_c)$ that separates the region with homogeneous flow from the region with shear band flow. The two values of $\dot{\gamma}_\mathrm{eff}$ in the heterogeneous regime have been measured separately in the two phases. The inset shows the width $w$ of the band as a function of the strain window $\Delta\gamma$, used for the averaging.}
  \label{fig-flowcurve2}
\end{figure}

Further we must check whether the mean field description is relevant for the spatial model. More precisely we have to test whether fluctuations of the effective strain rate are really irrelevant in the dynamics. We calculated numerically the steady state average stress as a function of the steady state average effective shear rate for a choice of the restructuring time for which we would expect the appearance of phase separation (see right panel of Figure \ref{fig-flowcurve2}). For large enough strain rates it turns out that the mean field model reproduces very well the dynamics of the spatial model. However for smaller strain rates there is a discrepancy between the two curves. This difference results from the stronger impact of the fluctuations of the effective shear rate on the dynamics, yielding a shift of the minimum of the flow curve towards smaller effective strain rates. This means that the mean field model underestimates the parameter range of homogeneous flow. But qualitatively we observe very similar dynamics as predicted by the mean field model. For effective strain rates below a critical value $\dot{\gamma}_\mathrm{eff}(\dot{\gamma}_c)$ the flow becomes heterogeneous with two effective strain rates. This accounts for the existence of a frozen phase with $\dot{\gamma}_\mathrm{eff}\approx 0$ and a strongly fluidized phase with $\dot{\gamma}_\mathrm{eff}\approx \dot{\gamma}_\mathrm{eff}(\dot{\gamma}_c)$. As shown in the inset of the right panel of Figure \ref{fig-flowcurve2} the effective strain rate is linear in $\dot{\gamma}$ for $\dot{\gamma}>\dot{\gamma}_c$ and the effective strain rate in the shear band region is independent of $\dot{\gamma}$ for $\dot{\gamma}<\dot{\gamma}_c$. Note that this scenario implies that the macroscopic flow curve, $\langle\sigma\rangle$ as a function of $\dot{\gamma}$, is a monotonic curve, not easily to be distinguished from a flow curve that corresponds to homogenous flow. Only the spatial analysis and the corresponding mean field study reveal the inhomogeneous nature of the flow.

In conclusion we can say that the mean field model produces qualitatively a good description of dynamics of the fully spatial model (see left panel of Fig.~\ref{fig-flowcurve2}. The qualitative behaviour of the two flow curves are the same. Both exhibit a minimum in the average stress as a function of shear rate for large restructuring times. This minimum explains the phase separation since for a given average stress, that is homogeneous in the system, two phases can coexist, one with zero effective shear rate and one with a finite effective shear rate. Hence, we provide a coherent scenario for the flow instability. Our results support strongly the idea that time-scales in the microscopic dynamics play an important role for the macroscopic flow behaviour.

Linking mesoscopic models such as the one studied above with 'real world' systems is a difficult issue.  One may proceed as Coussot and Ovarlez by infering from the consequences (namely the observed stability of bands) that some systems, such as attractive colloids, correspond to large $\tau_\mathrm{el}$, while other, such as foams, correspond to small $\tau_\mathrm{el}$. It is also interesting to speculate on the type of microscopic features that may give rise to a large restructuring time. Very generally, one can imagine that the restructuring time will be large in all cases where the shear can bring the system into a locally trapped configuration. The mechanical properties resulting from this metastable arrangement are thus expected to be significantly weaker than those of a relaxed configuration. Such a situation will be encountered for example in the case of strongly directional interactions (e.g. patchy colloids or covalently bonded systems \cite{Talati2009}). Another potentially interesting situation is the case where the interaction is mediated by a different component which may have a long relaxation time, as e.g. is the case for capillary bridges in some emulsions or in wet granular matter. One may also consider pair interactions between particles that involve an oscillatory potential and therefore a finite barrier between a 'bonded and a 'non bonded' state between neighboring particles. Interestingly, such a potential has been used for a long time to describe the formation of a single component quasicrystal \cite{dzugutov1993}, which is also a good glass former. It will be of interest to study the statistical properties of simple flows in such models in order to obtain a mapping onto our more coarse grained description.

\section*{Acknowledgements}
We acknowledge financial support from ANR, program SYSCOMM. KM is supported by 
the Marie Curie FP7-PEOPLE-2009-IEF program. JLB is supported by Institut Universitaire de France.



\footnotesize{
\providecommand*{\mcitethebibliography}{\thebibliography}
\csname @ifundefined\endcsname{endmcitethebibliography}
{\let\endmcitethebibliography\endthebibliography}{}

}

%
%
%
%

\begin{mcitethebibliography}{29}
\providecommand*{\natexlab}[1]{#1}
\providecommand*{\mciteSetBstSublistMode}[1]{}
\providecommand*{\mciteSetBstMaxWidthForm}[2]{}
\providecommand*{\mciteBstWouldAddEndPuncttrue}
  {\def\EndOfBibitem{\unskip.}}
\providecommand*{\mciteBstWouldAddEndPunctfalse}
  {\let\EndOfBibitem\relax}
\providecommand*{\mciteSetBstMidEndSepPunct}[3]{}
\providecommand*{\mciteSetBstSublistLabelBeginEnd}[3]{}
\providecommand*{\EndOfBibitem}{}
\mciteSetBstSublistMode{f}
\mciteSetBstMaxWidthForm{subitem}
{(\emph{\alph{mcitesubitemcount}})}
\mciteSetBstSublistLabelBeginEnd{\mcitemaxwidthsubitemform\space}
{\relax}{\relax}

\bibitem[Ragouilliaux \emph{et~al.}(2006)Ragouilliaux, Herzhaft, Bertrand, and
  Coussot]{Coussot2006}
A.~Ragouilliaux, B.~Herzhaft, F.~Bertrand and P.~Coussot, \emph{{R}heol
  {A}cta}, 2006, \textbf{46}, 261--271\relax
\mciteBstWouldAddEndPuncttrue
\mciteSetBstMidEndSepPunct{\mcitedefaultmidpunct}
{\mcitedefaultendpunct}{\mcitedefaultseppunct}\relax
\EndOfBibitem
\bibitem[Becu \emph{et~al.}(2006)Becu, Manneville, and Colin]{Becu2006}
L.~Becu, S.~Manneville and A.~Colin, \emph{{P}hys. {R}ev. {L}ett.}, 2006,
  \textbf{96}, 138302\relax
\mciteBstWouldAddEndPuncttrue
\mciteSetBstMidEndSepPunct{\mcitedefaultmidpunct}
{\mcitedefaultendpunct}{\mcitedefaultseppunct}\relax
\EndOfBibitem
\bibitem[Ragouilliaux \emph{et~al.}(2007)Ragouilliaux, Ovarlez,
  Shahidzadeh-Bonn, Herzhaft, Palermo, and Coussot]{Coussot2007}
A.~Ragouilliaux, G.~Ovarlez, N.~Shahidzadeh-Bonn, B.~Herzhaft, T.~Palermo and
  P.~Coussot, \emph{{P}hys. {R}ev. E}, 2007, \textbf{76}, 051408\relax
\mciteBstWouldAddEndPuncttrue
\mciteSetBstMidEndSepPunct{\mcitedefaultmidpunct}
{\mcitedefaultendpunct}{\mcitedefaultseppunct}\relax
\EndOfBibitem
\bibitem[Moller \emph{et~al.}(2009)Moller, Fall, Chikkadi, Derks, and
  Bonn]{Bonn2009}
P.~Moller, A.~Fall, V.~Chikkadi, D.~Derks and D.~Bonn, \emph{{P}hil. {T}rans.
  {R}. {S}oc. {A}}, 2009, \textbf{367}, 5139--5155\relax
\mciteBstWouldAddEndPuncttrue
\mciteSetBstMidEndSepPunct{\mcitedefaultmidpunct}
{\mcitedefaultendpunct}{\mcitedefaultseppunct}\relax
\EndOfBibitem
\bibitem[Divoux \emph{et~al.}(2011)Divoux, Tamarii, Barentin, and
  Manneville]{Manneville2011}
T.~Divoux, D.~Tamarii, C.~Barentin and S.~Manneville, \emph{arXiv:1110.1786},
  2011\relax
\mciteBstWouldAddEndPuncttrue
\mciteSetBstMidEndSepPunct{\mcitedefaultmidpunct}
{\mcitedefaultendpunct}{\mcitedefaultseppunct}\relax
\EndOfBibitem
\bibitem[Goyon \emph{et~al.}(2008)Goyon, Colin, Ovarlez, Ajdari, and
  Bocquet]{Goyon-Nature}
J.~Goyon, A.~Colin, G.~Ovarlez, A.~Ajdari and L.~Bocquet, \emph{{N}ature},
  2008, \textbf{454}, 84\relax
\mciteBstWouldAddEndPuncttrue
\mciteSetBstMidEndSepPunct{\mcitedefaultmidpunct}
{\mcitedefaultendpunct}{\mcitedefaultseppunct}\relax
\EndOfBibitem
\bibitem[Goyon \emph{et~al.}(2010)Goyon, Colin, and Bocquet]{Goyon-SoftMatter}
J.~Goyon, A.~Colin and L.~Bocquet, \emph{Soft Matter}, 2010, \textbf{6},
  2668\relax
\mciteBstWouldAddEndPuncttrue
\mciteSetBstMidEndSepPunct{\mcitedefaultmidpunct}
{\mcitedefaultendpunct}{\mcitedefaultseppunct}\relax
\EndOfBibitem
\bibitem[Fielding \emph{et~al.}(2009)Fielding, Cates, and
  Sollich]{SGR-shearbanding}
S.~M. Fielding, M.~E. Cates and P.~Sollich, \emph{Soft Matter}, 2009,
  \textbf{5}, 2378\relax
\mciteBstWouldAddEndPuncttrue
\mciteSetBstMidEndSepPunct{\mcitedefaultmidpunct}
{\mcitedefaultendpunct}{\mcitedefaultseppunct}\relax
\EndOfBibitem
\bibitem[Manning \emph{et~al.}(2007)Manning, Langer, and
  Carlson]{ManningLangerCarlson2007}
M.~L. Manning, J.~S. Langer and J.~M. Carlson, \emph{Phys. Rev. E}, 2007,
  \textbf{76}, 056106\relax
\mciteBstWouldAddEndPuncttrue
\mciteSetBstMidEndSepPunct{\mcitedefaultmidpunct}
{\mcitedefaultendpunct}{\mcitedefaultseppunct}\relax
\EndOfBibitem
\bibitem[Bocquet \emph{et~al.}(2009)Bocquet, Colin, and Ajdari]{KEP}
L.~Bocquet, A.~Colin and A.~Ajdari, \emph{{P}hys. {R}ev. {L}ett.}, 2009,
  \textbf{103}, 036001\relax
\mciteBstWouldAddEndPuncttrue
\mciteSetBstMidEndSepPunct{\mcitedefaultmidpunct}
{\mcitedefaultendpunct}{\mcitedefaultseppunct}\relax
\EndOfBibitem
\bibitem[Rodney \emph{et~al.}(2011)Rodney, Tanguy, and
  Vandembroucq]{Rodney2011}
D.~Rodney, A.~Tanguy and D.~Vandembroucq, \emph{arXiv:1107.2022}, 2011\relax
\mciteBstWouldAddEndPuncttrue
\mciteSetBstMidEndSepPunct{\mcitedefaultmidpunct}
{\mcitedefaultendpunct}{\mcitedefaultseppunct}\relax
\EndOfBibitem
\bibitem[Vandembroucq and Roux()]{Vandembroucq2011}
D.~Vandembroucq and S.~Roux, \emph{arXiv:1104.4863}\relax
\mciteBstWouldAddEndPuncttrue
\mciteSetBstMidEndSepPunct{\mcitedefaultmidpunct}
{\mcitedefaultendpunct}{\mcitedefaultseppunct}\relax
\EndOfBibitem
\bibitem[Moorcroft \emph{et~al.}(2011)Moorcroft, Cates, and
  Fielding]{Moorcroft2011}
R.~L. Moorcroft, M.~E. Cates and S.~M. Fielding, \emph{{P}hys. {R}ev. {L}ett.},
  2011, \textbf{106}, 055502\relax
\mciteBstWouldAddEndPuncttrue
\mciteSetBstMidEndSepPunct{\mcitedefaultmidpunct}
{\mcitedefaultendpunct}{\mcitedefaultseppunct}\relax
\EndOfBibitem
\bibitem[Jagla(2007)]{Jagla2007}
E.~A. Jagla, \emph{{P}hys. {R}ev. {E}.}, 2007, \textbf{76}, 046119\relax
\mciteBstWouldAddEndPuncttrue
\mciteSetBstMidEndSepPunct{\mcitedefaultmidpunct}
{\mcitedefaultendpunct}{\mcitedefaultseppunct}\relax
\EndOfBibitem
\bibitem[Jagla(2010)]{Jagla2010}
E.~A. Jagla, \emph{{J}. {S}tat. {M}ech.}, 2010,  P12025\relax
\mciteBstWouldAddEndPuncttrue
\mciteSetBstMidEndSepPunct{\mcitedefaultmidpunct}
{\mcitedefaultendpunct}{\mcitedefaultseppunct}\relax
\EndOfBibitem
\bibitem[Coussot and Ovarlez(2010)]{Coussot2010}
P.~Coussot and G.~Ovarlez, \emph{{E}ur. {P}hys. {J}. {E}.}, 2010, \textbf{33},
  183--188\relax
\mciteBstWouldAddEndPuncttrue
\mciteSetBstMidEndSepPunct{\mcitedefaultmidpunct}
{\mcitedefaultendpunct}{\mcitedefaultseppunct}\relax
\EndOfBibitem
\bibitem[Dahmen \emph{et~al.}(2009)Dahmen, Ben-Zion, and Uhl]{Dahmen2009}
K.~A. Dahmen, Y.~Ben-Zion and J.~T. Uhl, \emph{Phys. Rev. Lett.}, 2009,
  \textbf{102}, 175501\relax
\mciteBstWouldAddEndPuncttrue
\mciteSetBstMidEndSepPunct{\mcitedefaultmidpunct}
{\mcitedefaultendpunct}{\mcitedefaultseppunct}\relax
\EndOfBibitem
\bibitem[Mansard \emph{et~al.}(2011)Mansard, Colin, Chauduri, and
  Bocquet]{Mansard2011}
V.~Mansard, A.~Colin, P.~Chauduri and L.~Bocquet, \emph{{S}oft {M}atter}, 2011,
  \textbf{7}, 5524--5527\relax
\mciteBstWouldAddEndPuncttrue
\mciteSetBstMidEndSepPunct{\mcitedefaultmidpunct}
{\mcitedefaultendpunct}{\mcitedefaultseppunct}\relax
\EndOfBibitem
\bibitem[Martens \emph{et~al.}(2011)Martens, Bocquet, and Barrat]{MBB-PRL2011}
K.~Martens, L.~Bocquet and J.-L. Barrat, \emph{{P}hys. {R}ev. {L}ett.}, 2011,
  \textbf{106}, 156001\relax
\mciteBstWouldAddEndPuncttrue
\mciteSetBstMidEndSepPunct{\mcitedefaultmidpunct}
{\mcitedefaultendpunct}{\mcitedefaultseppunct}\relax
\EndOfBibitem
\bibitem[Barrat and Lema\^itre(2011)]{BarratLemaitre2011}
J.-L. Barrat and A.~Lema\^itre, \emph{in Dynamical Heterogeneities in glasses,
  colloids and granular materials, L. Berthier et al Eds ; arXiv:1009.5774v3},
  Oxford University Press, 2011, ch. Heterogeneities in amorphous systems under
  shear\relax
\mciteBstWouldAddEndPuncttrue
\mciteSetBstMidEndSepPunct{\mcitedefaultmidpunct}
{\mcitedefaultendpunct}{\mcitedefaultseppunct}\relax
\EndOfBibitem
\bibitem[Picard \emph{et~al.}(2004)Picard, Ajdari, Lequeux, and
  Bocquet]{Picard2004}
G.~Picard, A.~Ajdari, F.~Lequeux and L.~Bocquet, \emph{{E}ur. {P}hys. {J}.
  {E}}, 2004, \textbf{15}, 371\relax
\mciteBstWouldAddEndPuncttrue
\mciteSetBstMidEndSepPunct{\mcitedefaultmidpunct}
{\mcitedefaultendpunct}{\mcitedefaultseppunct}\relax
\EndOfBibitem
\bibitem[Picard \emph{et~al.}(2005)Picard, Ajdari, Lequeux, and
  Bocquet]{Picard2005}
G.~Picard, A.~Ajdari, F.~Lequeux and L.~Bocquet, \emph{{P}hys. {R}ev. {E}},
  2005, \textbf{71}, 010501(R)\relax
\mciteBstWouldAddEndPuncttrue
\mciteSetBstMidEndSepPunct{\mcitedefaultmidpunct}
{\mcitedefaultendpunct}{\mcitedefaultseppunct}\relax
\EndOfBibitem
\bibitem[Argon(1979)]{Argon1979}
A.~Argon, \emph{Acta Met}, 1979, \textbf{27}, 47\relax
\mciteBstWouldAddEndPuncttrue
\mciteSetBstMidEndSepPunct{\mcitedefaultmidpunct}
{\mcitedefaultendpunct}{\mcitedefaultseppunct}\relax
\EndOfBibitem
\bibitem[Tanguy \emph{et~al.}(2006)Tanguy, Leonforte, and Barrat]{Tanguy2006}
A.~Tanguy, F.~Leonforte and J.~Barrat, \emph{{E}ur. {P}hys. {J}. {E}}, 2006,
  \textbf{20}, 355\relax
\mciteBstWouldAddEndPuncttrue
\mciteSetBstMidEndSepPunct{\mcitedefaultmidpunct}
{\mcitedefaultendpunct}{\mcitedefaultseppunct}\relax
\EndOfBibitem
\bibitem[Schall \emph{et~al.}(2007)Schall, Weitz, and Spaepen]{Schall2007}
P.~Schall, D.~Weitz and F.~Spaepen, \emph{Science}, 2007, \textbf{318},
  1895\relax
\mciteBstWouldAddEndPuncttrue
\mciteSetBstMidEndSepPunct{\mcitedefaultmidpunct}
{\mcitedefaultendpunct}{\mcitedefaultseppunct}\relax
\EndOfBibitem
\bibitem[Eshelby(1957)]{Eshelby}
J.~Eshelby, \emph{Proc. R. Soc. London, Ser. A}, 1957, \textbf{241}, 467\relax
\mciteBstWouldAddEndPuncttrue
\mciteSetBstMidEndSepPunct{\mcitedefaultmidpunct}
{\mcitedefaultendpunct}{\mcitedefaultseppunct}\relax
\EndOfBibitem
\bibitem[Sollich \emph{et~al.}(1997)Sollich, Lequeux, Hebraud, and
  Cates]{SGR97}
P.~Sollich, F.~Lequeux, P.~Hebraud and M.~Cates, \emph{Phys. Rev. Lett.}, 1997,
  \textbf{78}, 2020\relax
\mciteBstWouldAddEndPuncttrue
\mciteSetBstMidEndSepPunct{\mcitedefaultmidpunct}
{\mcitedefaultendpunct}{\mcitedefaultseppunct}\relax
\EndOfBibitem
\bibitem[Talati \emph{et~al.}({2009})Talati, Albaret, and Tanguy]{Talati2009}
M.~Talati, T.~Albaret and A.~Tanguy, \emph{{EPL}}, {2009}, \textbf{{86}},
  66005\relax
\mciteBstWouldAddEndPuncttrue
\mciteSetBstMidEndSepPunct{\mcitedefaultmidpunct}
{\mcitedefaultendpunct}{\mcitedefaultseppunct}\relax
\EndOfBibitem
\bibitem[Dzugutov(1993)]{dzugutov1993}
M.~Dzugutov, \emph{Phys. Rev. Lett.}, 1993, \textbf{70}, 2924--2927\relax
\mciteBstWouldAddEndPuncttrue
\mciteSetBstMidEndSepPunct{\mcitedefaultmidpunct}
{\mcitedefaultendpunct}{\mcitedefaultseppunct}\relax
\EndOfBibitem
\end{mcitethebibliography}

\end{document}